# Electron Lenses, Tevatron and Selected Topics in Accelerators : 2019 Nishikawa Prize Talk

*Vladimir Shiltsev*

*Fermi National Accelerator Laboratory, PO Box 500, Batavia, IL 60510, USA*

*Abstract*

This article is an extended version of the talk is given at the IPAC19 (Melbourne, Australia, May 2019) on the occasion of acceptance of the ACFA/IPAC19 Nishikawa Tetsuji Prize for a recent, significant, original contribution to the accelerator field, with no age limit with citation "*...for original work on electron lenses in synchrotron colliders, outstanding contribution to the construction and operation of high-energy, high-luminosity hadron colliders and for tireless leadership in the accelerator community.*"

PACS numbers: 29.27.Bd, 29.20.dk, 52.35.Qz, 41.85.Ew

## 1. TRIBUTE TO NISHIKAWA TESUJI

It is a great honour for me to receive the 2019 ACFA/IPAC Nishikawa Tetsuji Prize. Prof. Nishikawa – see Fig.1 - was among the pioneers of particle accelerators who shaped our field at its early stage. After making a significant contribution to the BNL linac in 1964-1966, he moved to Japan and established Japan National Lab for High Energy Physics in 1969 (now KEK). There he initiated design, construction, commissioning and operation of such remarkable facilities as the 12 GeV proton synchrotron[1], neutron facility J-PARC, the 500 MeV cancer treatment synchrotron, the KEK Photon Factory, and TRISTAN electron-positron collider. Prof. Nishikawa was also a man of great integrity and an active supporter of the international cooperation, in particular, between Japan and the US [2].

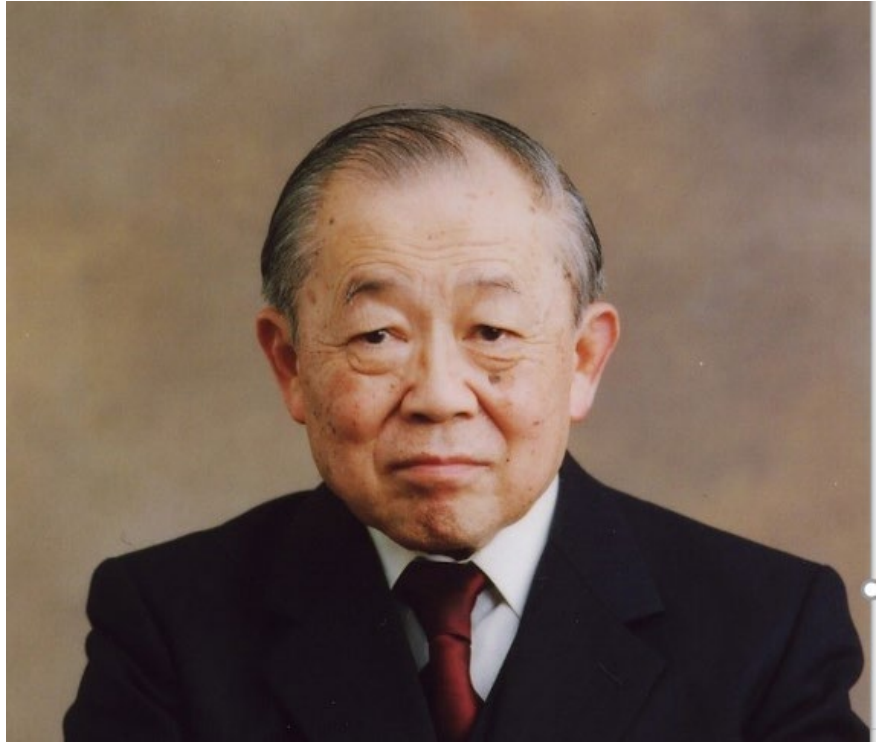

Figure 1: Tetsuji Nishikawa (1926-2010).

In this article I summarize the activities of myself and my teams noted in the prize citation and briefly discuss the topics of my current research.

## 2. ELECTRON LENSES

I got a PhD in accelerator and beam physics from BINP (Novosibirsk, Russia) in 1994, and worked at the leading accelerator laboratories in Protvino (Russia), the SSCL (USA) and DESY (Germany) before joining Fermilab in 1996 where shortly thereafter I initiated the project of beam-beam compensation with *Tevatron Electron Lenses* (TELs) [3,4,5]. Since then the electron lenses - a novel instrument for high-energy particle accelerators – have been added to the toolbox of modern beam facilities, being particularly useful for the energy frontier superconducting hadron colliders ("supercolliders") [6]. The physics mechanism of the electron lens is the space-force of a narrow and long beam of low energy electrons, immersed in strong longitudinal magnetic field. The Lorentz force of a low energy electron beam with velocity $\beta_e c$ and current density distribution $j_e(r)$ is

$$e(E_r + B_\theta) = \frac{4\pi e(1+\beta_e)}{\beta_e c} \frac{1}{r} \int_0^r j_e(r') r' dr' . \tag{1}$$

It effectively acts on the accelerator's high energy beam moving along (or colliding) with the electron beam while the effect of the longitudinal magnetic field is usually a minor or easily correctable imperfection. The transverse motion of electrons is essentially frozen along the magnetic field lines which, therefore, assures outstanding stability of the electron beam. Given that the electron beam transverse shapes and longitudinal current modulation patterns can be broadly varied (usually, created at an electron gun) [7] the electron lenses have become a uniquely flexible instrument.

The electron lenses for supercolliders were originally considered in 1993 by Tsyganov *et al* [8] to reduce the tune spread due to beam-beam interactions in the SSC and, independently, in 1997 Shiltsev *et al* proposed them for compensation of the long-range and head-on beam-beam effects in the Tevatron proton-antiproton collider [3, 4]. Comprehensive theory of the electron lens beam-beam compensation and detail design considerations were accomplished by 1999 [5]. Then, at Fermilab, the first two TELs were designed, built and installed in the Tevatron 1.96 TeV c.m.e. proton-antiproton collider in 2001 and 2004 (see Figure 2), operated till the end of the Run II in 2001 and were used for compensation of long-range beam-beam effects (the TELs varied tune shift of selected individual 0.98 TeV antiproton or proton bunches by 0.003-0.01) [9] – see Figure 3; longitudinal collimation – removal the DC beam particles from the abort gaps - for 10 years in regular operation [10]; studies of head-on beam-beam compensation [11,12]; demonstration of halo scraping with hollow electron beams [13, 14]. Since 2015 two electron lenses are installed in RHIC at BNL and very successfully used for head-on beam-beam compensation leading to doubling the luminosity in proton-proton collisions [15].

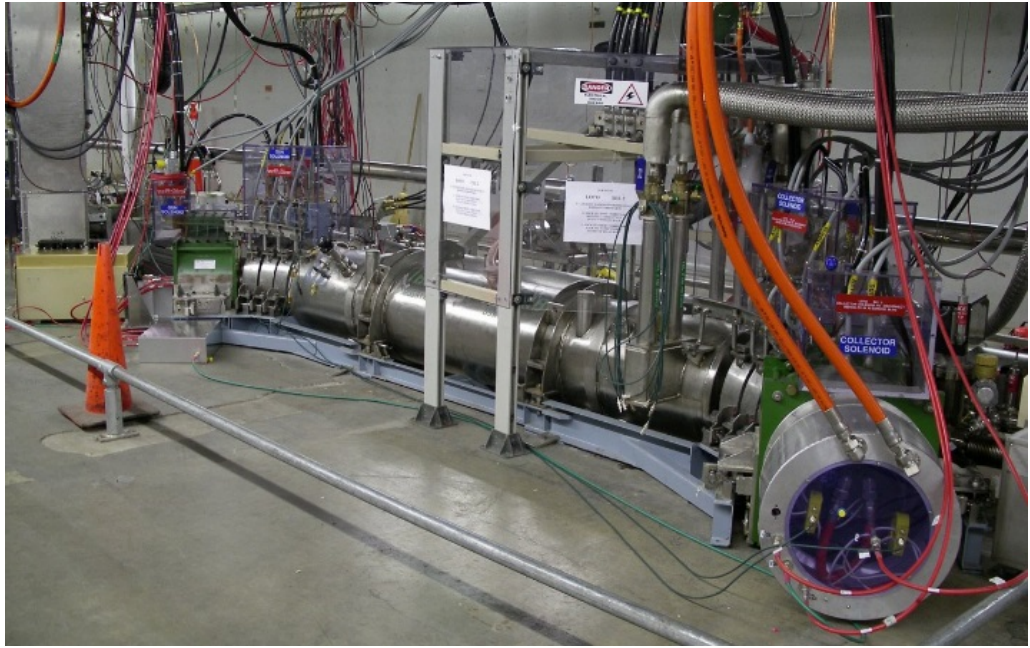

Figure 2: Tevatron electron lens in the accelerator tunnel.

Present world-wide efforts on the electron lenses cover several areas of research: a) hollow electron beam collimation of protons in the HL-LHC [16, 17]; b) long-range beam-beam compensation with electron lenses as current-bearing "wires" in the HL-LHC [18-20]; c) generation of nonlinear integrable lattices with special transverse current distribution $j_e(r) \sim /(1+r^2)^2$ – first proposed in [21, 5] - eg in the IOTA ring [22, 23]; d) Landau damping of broad spectrum of coherent instabilities by the electron lens induced tune spread [13, 24] in, e.g., the LHC, FCC-hh (where an electron lens can outperform some 10,000 octupoles), or FNAL Recycler; e) compensation of space-charge effects in modern high-intensity rapid cycling synchrotrons (RCS) [25, 26].

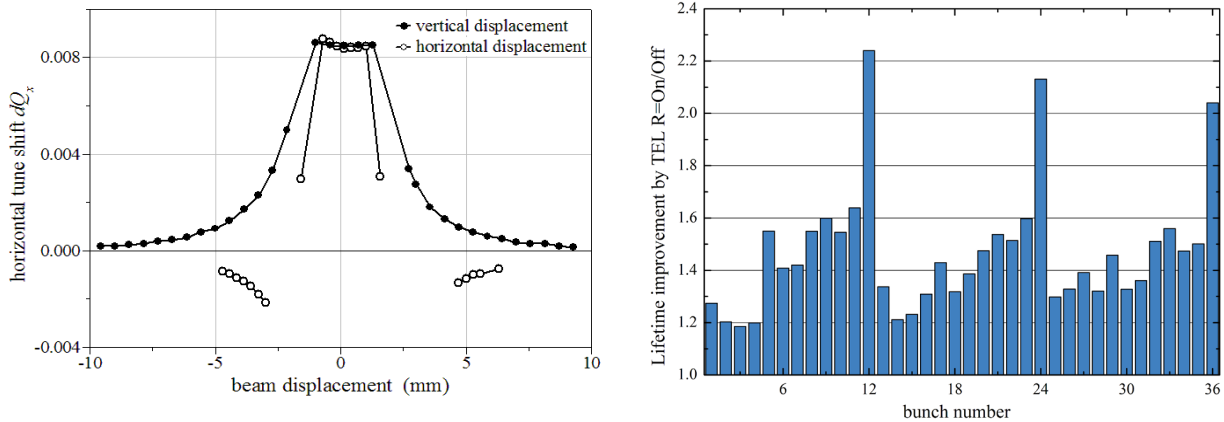

Figure 3 : a) left - TEL1-induced shift of 980 GeV proton horizontal betatron tune versus vertical (filled circles) and horizontal (open circles) electron beam displacement ($J_e$ =1A, $U_c$ =6.0 kV, electron gun with uniform transverse current distribution); b) TEL-2 in DC regime significantly improves the lifetime of all 36 proton bunches early in the Tevatron store #5183 with initial luminosity $L$=253×$10^{30}$ cm$^{-2}$s$^{-1}$ – from Ref. [11].

The electron lenses usually employ low energy (~10kV), high current (1-10 A) mm or sub-mm size magnetized electron beam to affect beneficially a high energy beam of hadrons (protons, antiprotons). The design of the lenses required advancing several technologies: high field quality solenoids and correctors (sometimes superconducting, with up to 4-6 T field), high brightness electron beam generation and low loss transport (gun, collector, etc), fast HV gun anode modulator (10kV, 100's μs and multiples of the machine revolution frequency, e.g. 50 kHz in the

Tevatron), sophisticated power recirculation electrical scheme, ultra-high vacuum and beam diagnostics system.

Together with other beam physics R&D, further research on the electron lenses motivated the construction of a dedicated accelerator test facility –Integrable Optics Beam Accelerator (IOTA) ring at Fermilab [27]. The 40-m circumference ring, capable operateing with 70-150 MeV/c electrons and protons – see Fig.4 - has been commissioned in 2018 and the first experimental run with electrons has taken place in Oct. 2018 – Mar. 2019, providing many excellent results in key experiments on the nonlinear integrable optics, single-electron tomography and many others – see, e.g., related reports at the IPAC'19 [28-45]. The experiment on space-charge compensation with an electron lens is being prepared with the goal to be ready for the IOTA operation with protons in 2020-2021.

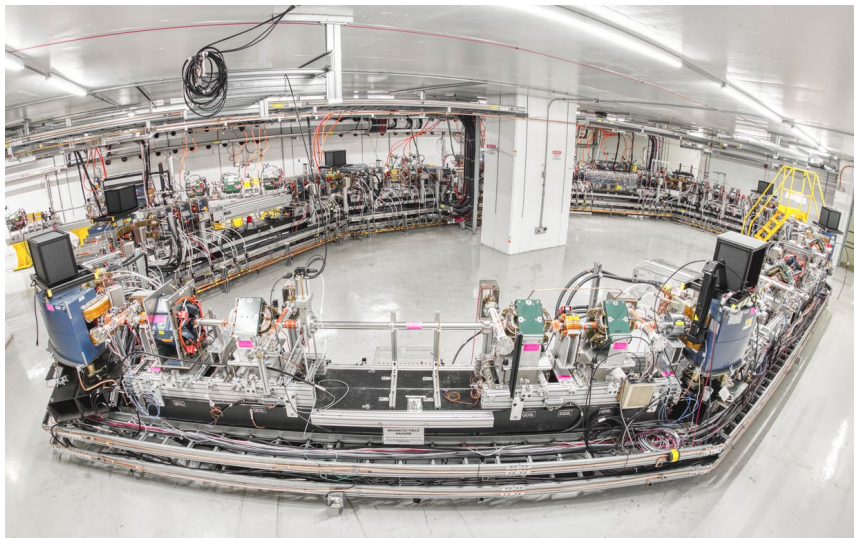

Figure 4: Fermilab's IOTA ring for beam physics research.

## 3. TEVATRON COLLIDER IMPROVEMENTS

Over more than three decades in accelerators, I have been involved in, contributed to or led design, construction or operation of a number of forefront accelerators and colliders, such as Tevatron, VLEPP, TESLA, UNK, HERA, PIP-II, LHC, Muon Collider, ILC, etc. The most extensive work was done for the Tevatron collider which took over a decade [46-49]. Through computer modeling and a series of dedicated beam studies we greatly advanced the understanding of a number of critical beam physics phenomena creating impediments on the collider's luminosity, such as long-range and head-on beam-beam effects [49], beam injection optics matching, control of coherent beam instabilities, two-stage halo collimation, novel collimation methods by hollow electron beams collimation method [12] and bent crystals [50]. The beam-beam effects significantly affected the collider operation and resulted in either emittance growth or bunch length reduction or particle losses at all stages from injection to energy ramp and squeeze as well as in collisions – see Fig.5 – and via series of dedicated studies and analysis of operational data we were able to establish functional dependencies of the effects on such key parameters as the opposite beam intensity and emittance, on the beam's own emittance, on beam-beam separation, machine tunes, chromaticities, etc.

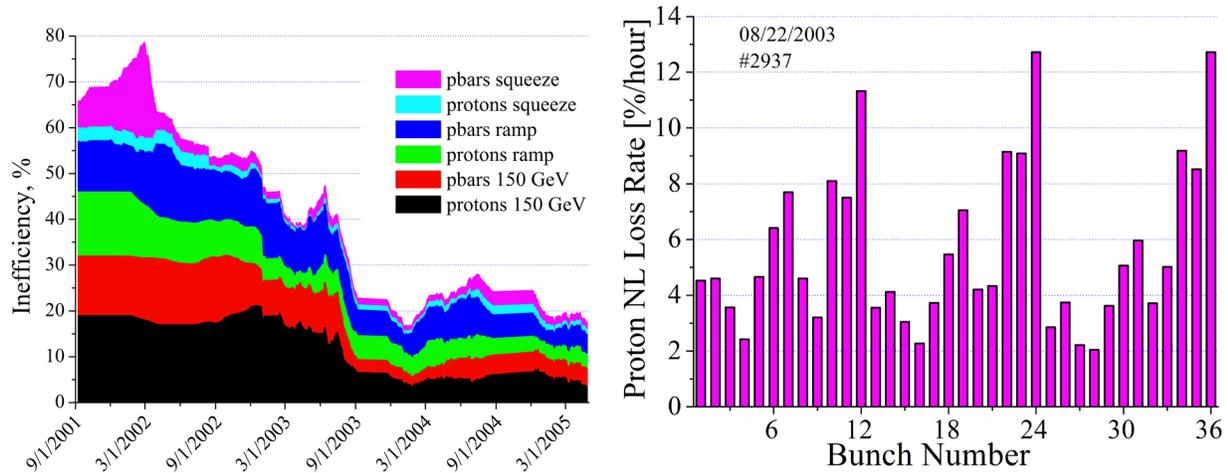

Figure 5: a) left - improvements of the Tevatron operational inefficiencies (ratios of the total bunch intensities after and before various stage) in 2001-2005; b) right - non-luminous loss rate of proton bunch intensity at the beginning of store #2973 (August 22, 2003) – from Ref.[49].

The first ever two-stage beam halo collimation system using bent Si crystals was installed, tested and introduced into operation of the Tevatron collider where it has demonstrated the predicted excellent performance [51]. Since then, bent crystal systems were developed and installed in the LHC where they demonstrated excellent performance as well. Beam diagnostics was greatly improved over the collider Run II years of 2001-2011 (Schottky detectors, ionization profile monitors, new BPMs, etc) [52].

# 4. ATL LAW OF DIFFUSIVE GROUND MOTION

Stability of beam orbits in large colliders is often affected by the *ATL*-type of the ground diffusion of accelerator tunnels [53]. The *ATL* law approximates the variance $<dY^2(L)>=<(dY(z)-dY(z+L))^2>$ of the elevation difference of the points of a tunnel as a function of the lag $L$ (distance between pairs of the measurement points) averaged over all possible time intervals $T$ as :

$$<dY^2>=ATL, \qquad (2)$$

where $A$ is a site dependent diffusion constant [54]. More than two decades of vibrations and ground motion studies [55] for accelerator tunnels, development of advanced instruments [56] resulted in solid experimental confirmation of *the ATL law* and comprehensive understanding of its practical implications for operation of all large accelerators. In particular, the Tevatron alignment "tie rod" data presented in Fig.6 give the ground diffusion coefficient $A=(4.9\pm0.13)\cdot10^{-6}$ µm$^2$/s/m [57]. Corresponding diffusive orbit deterioration due to the ground motion was clearly observed in the Tevatron, LEP, SPS, KEK-B and TRISTAN [54].

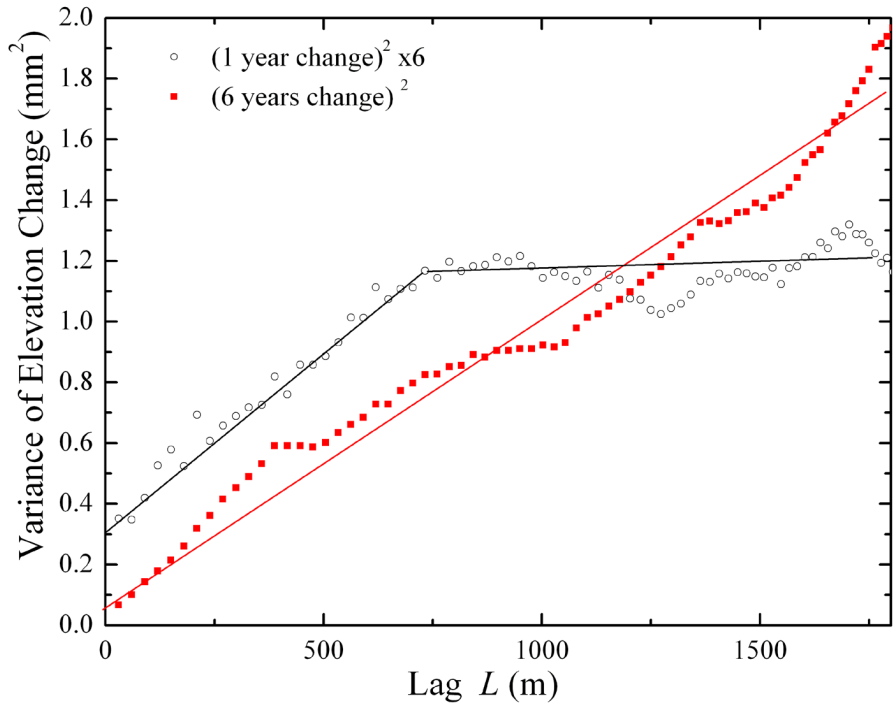

Figure 6: Variances of the Tevatron alignment "tie rod" vertical displacements over time intervals of 1 year (multiplied by 6) and 6 years vs the distance $L$ [57]. Solid lines are according to the *ATL law*.

## 5. CPT THEOREM OF ACCELERATOR COMMISSIONING

The luminosity upgrade of the Tevatron proton-antiproton collider and corresponding beam physics tests and studies resulted in a ~40 fold improvement of then world's most powerful accelerator [58]. Accelerator physics and technology advances implemented over the course of the Tevatron operation have been comprehensively summarized in our book [59] and were of big help during beam commissioning of the LHC at CERN. Analysis of the luminosity evolution of the Tevatron which occurred as a sequence of medium- to small-scale improvements, augmented with similar observations at other accelerators was summarized in a *CPT-theorem* for accelerators [60] that observes that in the case of the typical progress like "N% gain per step, step after step, with regular periodicity" explains the typical exponential growth of the luminosity over time interval $T$ as $L(t_0 + T) = L(t_0) \times e^{T/C}$, or

$$CP=T, \qquad (3)$$

where $P=ln(L(t_0 + T)/L(t_0))$ is the performance progress and $C$, the complexity coefficient, usually varies between 1 year to 3 years depending on the type of accelerator.

## 5. $\alpha\beta\gamma$-MODEL OF COST OF LARGE ACCELERATORS

In a further analysis of various large accelerators and colliders I turned to their costs and found out that based on publicly available costs for 17 large accelerators of the past, the present and those currently in the planning stage, the "total project cost (TPC)" - sometimes cited as "the US accounting" - of a collider can be broken up into three major parts corresponding to "civil construction", "accelerator components", "site power infrastructure". The three respective cost components can be parameterized by just three parameters – the total length of the facility tunnels $L_f$, the center-of-mass or beam energy $E$, and the total required site power $P$ - and over almost 3 orders of magnitude of $L_f$, 4.5 orders of magnitude of $E$ and more than 2 orders of magnitude of $P$ the so-called "$\alpha\beta\gamma$-cost model" works with ~30% accuracy [61]:

$$TPC \approx \alpha \times (Length/10km)^{1/2} + \beta \times (Energy/TeV)^{1/2} + \gamma \times (Power/100MW)^{1/2}, \quad (4)$$

where coefficients $\alpha=2B\$/(10\ km)^{1/2}$, $\gamma=2B\$/(100MW)^{1/2}$, while the accelerator technology dependent coefficient $\beta$ is equal to $10\ B\$/TeV^{1/2}$ for the superconducting RF accelerators, $8\ B\$/TeV^{1/2}$ for normal-conducting ("warm") RF, $1B\$/TeV^{1/2}$ for normal-conducting magnets and $2B\$/TeV^{1/2}$ for SC magnets (all numbers in 2014 US dollars). This scaling law was found extremely useful for ball-park cost estimates of possible future accelerators considered in the course of the US and European particle physics strategy discussions in 2013-2019.

## 5. MUON COLLIDERS

Already during the Tevatron years I got involved in the design studies, experimental tests and leadership of the Fermilab Muon Collider Task Force and later, in the US Muon Accelerator Program. These studies were focused on the muon ionization cooling, NC RF in strong magnetic fields and overall machine design and established the accelerator physics feasibility of the muon collider – an alternative cost- and power-efficient energy frontier machine for the future particle physics research [63, 64]. Now the muon colliders are under active consideration for the European particle physics strategy, with especially attractive concept of a 14 TeV c.m.e. $\mu^+\mu^-$ collider in the LHC tunnel [64]. Related activity was the development of the HTS based fast cycling magnets which could offer economical way to build rapid cycling synchrotrons for particle physics (e.g., proton sources and muon accelerators), neutron spallation sources and other applications [65, 66]. In 2018 we have achieved the magnetic field ramping rate of 12 T/s in a dual bore HTS magnet prototype, exceeding the previous world record rate for SC magnets by a factor of 3 [67].

## 6. RESEARCH AT IOTA RING AND ELSEWHERE

As part of the FNAL Accelerator Physics Center (APC) mission (see below), since 2009 we constructed the FAST facility which beside the IOTA ring comprises of a 300 MeV 1.3GHz SC RF electron linac [27]. In the Fall of 2017, we achieved the world record-high superconducting RF beam accelerating gradient of 31.5 MeV/m

[68] - this was the first ever demonstration of average beam accelerating gradient matching the specifications of the International Linear Collider (ILC). That accomplishment greatly boosted the confidence in the technical feasibility of the ILC – the supercollider project to push elementary particle physics beyond LHC and which is now in the final stage of approval by the Japanese government. The FAST linac was also commissioned as an electron injector to IOTA ring and already supports user research [69]. Of note here is that my involvement in the linear colliders' R&D began in 1990's back in Russia and at DESY and, e.g. I was part of the group which for the first time demonstrated ns-scale HV pulse (~5kV, high rep rate) travelling wave injection/extraction kickers for multi-bunch storage rings and other accelerator applications [70], thus, paving the way to many modern pulsed ns HV systems based on MOSFETs, FIDs and other fast switches and allowing to shorten the circumference of the damping rings of the future *e+e-* linear colliders from 10-20 km to about 3 km. Also at DESY, I came out with the idea of even faster beam-beam kicker [71] and co-authored a seminal paper on the theory of coherent synchrotron radiation [72]. In 2005-2007 I led the Accelerator Systems section of the US LARP (LHC Accelerator Research Program) and either supervised and participated in or led the beam physics analysis, technical design, construction or operation of such systems as high frequency Schottky detectors, rotatable collimators, simulations of the beam-beam effects and their compensation by electron

lenses, etc. All these studies helped the LHC and its upgrades as well as future proton supercolliders [73].

## 7. ACCELERATOR PHYSICS CENTER

Accelerator construction, operation and research by necessity involve big teams of physicists, engineers and technicians. Over more than 20 years at Fermilab I led the beam-beam compensation project group (1997-2001), then I was the Head of the Tevatron Department in 2001-2006, and was appointed the inaugural Director of the Fermilab Accelerator Physics Center (APC, 2007-2018) [74]. APC was a unique organization created in June 2007 with mission to carry out R&D to keep the US leading high-energy physics laboratory at the forefront of accelerator science, technology and facility operation. In support of the FNAL high-energy physics research mission, APC scientists and engineers conducted accelerator R&D aimed at next-generation and beyond accelerator facilities; provided accelerator physics support for existing operational programs and the evolution thereof; trained accelerator scientists and engineer and established experimental programs for a broad range of accelerator R&D that can be accessed by both Fermilab staff and by the US and world HEP community. APC was a center-place for in-depth design, research and development efforts which allowed the Laboratory to make intelligent decisions on the ILC in the US and on the Muon Collider, as well as to initiate projects such as PIP-II (through the Proton Driver/Project-X work), LHC-AUP (via LARP) and the IOTA/FAST R&D facility. The results of the APC researchers were published in 215 peer-review articles and were used as an input for the discussions of the Fermilab Steering Group, of the P5 (Particle Physics Project Prioritization Panel) in 2008, during the 2014 HEPAP P5 process and the 2015 HEPAP Accelerator R&D Subpanel meetings. Important contributions were made to the ILC CRD and TDR, to the Muon Collider design studies - see the JINST collection of reports and papers on muon accelerators [75], as well as the Project-X CDR [76] and the PIP-II CDR [77], the High-Luminosity LHC Upgrade project [78] and the IOTA/FAST R&D facility CDR [27].

Experimental beam physics R&D was carried out at the operational accelerators at Fermilab and CERN as well as at a number of dedicated beam test facilities developed, constructed or supported by APC for the purpose of accelerator R&D: the Fermilab NICADD Photoinjector Laboratory with an 18 MeV electron linac; the MuCool Test Area with a 201-MHz test cavity installed next to 5 T solenoid, also used for testing the smaller 805-MHz cavities under the impact of a 400 MeV proton beam; the Muon Ionization Cooling Experiment (MICE) at RAL (UK) which operated with 200 MeV muons; 325 MHz HINS RFQ that accelerated protons and H- particles to 2.5 MeV; 1.3 GHz SRF 300 MeV electron injector at FAST and IOTA ring operating with up to 150 MeV/c electrons and 70 MeV/c protons.

. APC also was the birthplace and host of many national and international collaborations and several educational and training programs in beam physics resulted in 27 PhD theses. In addition to the office of US Particle Accelerator School (US PAS), APC hosted several programs in accelerator and beam physics education and training such as the Lee Teng Internship (jointly with ANL), the Helen Edwards Internship for international students, and Joint Fermilab-University PhD Program in accelerator physics. APC hosted 5 Joint Appointments with NIU and IIT, as well 11 Peoples Fellows, 3 US LARP Toohig Fellows and 2 Imperial College London Fellows. Two out of the 27 students graduated from the PhD program won the APS DPB Outstanding Doctoral Thesis Research in Beam Physics Award. In 2007-2018, APC hosted more than 100 visitors and PhD or MSc students, over 100 summer students in the Lee Teng and Helen Edwards programs, and some 3000 students attended semi-annual US PAS sessions.

## 8. NEXT STEPS : EXCITING OPPORTUNITIES

At present, I am very interested in and participate in two research projects: the first is modelling of the space-charge compensation (SCC) with electron lenses [25] in space-charge dominated ultimate-intensity synchrotrons with $dQ_{SC} \sim -1.0$ (jointly with E.Stern, A.Burov and Yu.Alexahin). The simulation results show great promise of an about order of magnitude reduction in the beam losses with electron lenses – see Fig.7 [79]. The ultimate goal of the study is to understand the requirements on, and key specifications of the electron lenses for the SCC and to evaluate their effectiveness for record beam power RCS design, as well as to make verifiable predictions for the space-charge compensation demonstration experiment in IOTA. The other project is an initial exploration of the ultimate high gradient acceleration of muons in crystals and nanostructures – the idea that may potentially open the path to PeV class colliders [80]. In collaboration with the pioneer of the method, Prof. T.Tajima, we organize a "Workshop on Beam Acceleration in Crystals and Nanostructures" this year at Fermilab [81].

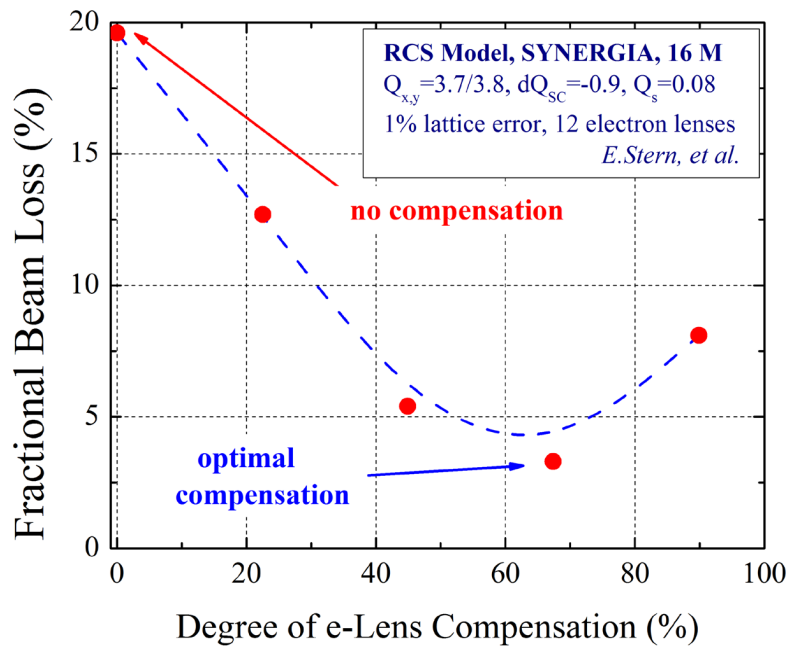

Figure 7: Fractional loss of proton beam with $dQ_{SC}$ =-0.9 after 1000 turns in a model RCS with 12 electron lenses vs the degree of the SC compensation (PIC simulations data courtesy E. Stern).

## ACKNOWLEDGEMENTS

I would like wholeheartedly thank those who nominated me and many colleagues I had fortune to work with over many years on the electron lenses, the Tevatron collider and many interesting and important topics from beam-beam effects to bent crystal collimation, ground motion and orbit stabilization, head-tail instability and super-fast HV pulsers, future collider designs and construction of IOTA ring, beam commissioning of the worlds' best ILC cryo-module and on very fast cycling HTS magnets. I'd like to pay special tribute to my late collaborators D. Wildman (FNAL), G.F. Kuznetsov (BINP/FNAL), M. Tiunov (BINP) and V. Danilov (ORNL).

Important contributions to the conceptual and technical development of the electron lens method were made by D. Finley, A. Valishev, G. Stancari, N. Solyak, A. Burov, E. Stern and Yu. Alexahin (FNAL), V. Parkhomchuk and D. Shatilov (BINP), E. Tsyganov (JINR/SSCL), A. Seryi (JLab), K. Bishofberger (LANL), V.Kamerdzhiev (COSY), W. Fischer (BNL), S. Redaelli and F. Zimmermann (CERN) and I greatly appreciate many years of fruitful collaboration with them.

Fermilab is supported by U.S. Department of Energy, Office of Science, Office of High Energy Physics, under Contract No. DE-AC02-07CH11359.